\documentclass[a4paper,11pt]{article}

\usepackage[margin=1in]{geometry}

\usepackage[english]{babel}
\usepackage{amssymb}
\usepackage{amsmath}
\usepackage{amsthm}
\usepackage{psfrag}
\usepackage[T1]{fontenc}
\usepackage{ae,aecompl}
\usepackage[colorlinks]{hyperref}
\usepackage{subfigure}
\usepackage{appendix}
\usepackage{natbib}
\hypersetup{colorlinks,breaklinks=true,citecolor=blue,linkcolor=blue}
\usepackage{graphicx}
\usepackage{color}

\newcommand{\Pm}{\textrm{Pm}}
\newcommand{\Pran}{\textrm{Pr}}
\newcommand{\Ek}{\textrm{Ek}}
\newcommand{\Ra}{\textrm{Ra}}
\newcommand{\tRa}{\widetilde{\textrm{Ra}}}

\newcommand{\Rm}{\mbox{\textrm{Rm}}}
\newcommand{\Rey}{\mbox{\textrm{Re}}}
\newcommand{\aspect}{\lambda}
\newcommand{\pleft}{\left(}
\newcommand{\pright}{\right)}
\newcommand{\B}{\boldsymbol{B}}
\newcommand{\vel}{\boldsymbol{u}}

\newcommand{\pdt}[1]{\frac{\partial #1}{\partial t}}
\newcommand{\vect}[1]{\boldsymbol{#1}}
\providecommand\bnabla{\boldsymbol{\nabla}}
\providecommand\bcdot{\boldsymbol{\cdot}}

\newcommand\eg{e.g.\ }
\newcommand\ie{i.e.\ }

\title{Generation of magnetic fields by large-scale vortices in rotating convection}

\author{C\'eline Guervilly, David W. Hughes \& Chris A. Jones \vspace{0.2cm}
 \\ {\small Department of Applied Mathematics, University of Leeds, Leeds LS2 9JT, UK}}
 
\begin{document}

\maketitle

\begin{abstract}
We propose a new self-consistent dynamo mechanism for the generation of large-scale magnetic fields in natural objects. Recent computational studies have described the formation of large-scale vortices (LSVs) in rotating turbulent convection. Here we demonstrate that for magnetic Reynolds numbers below the threshold for small-scale dynamo action, such turbulent
flows can sustain large-scale magnetic fields --- i.e.\  fields with a significant component on the scale of the system.
\end{abstract}

Many astrophysical bodies 
possess large-scale magnetic fields. These are believed to be the products of hydromagnetic dynamo action, in which the inductive motions of an electrically conducting fluid, typically driven by thermal convection, maintain the magnetic field against Ohmic dissipation. In rapidly rotating, low-viscosity astrophysical bodies, convective flows appear on scales small compared with the system size. The most important and long-standing question in dynamo theory thus concerns the mechanism by which such small-scale flows can produce magnetic fields at large scales (\ie of a size comparable with that of the body itself). 
In this paper we demonstrate a new mechanism for the generation of large-scale magnetic fields, based on the formation of large-scale vortices in rotating turbulent convection.

Significant progress on the problem of large-scale magnetic field generation has been achieved through
computational models, which have
shown that convective flows can indeed produce large-scale fields \citep[\eg][]{Stellmach2004, Christensen2011}. In these models, the generation of magnetic fields with a pronounced large-scale component relies on the presence of coherent 
convective vortices aligned with the rotation axis \citep{Childress1972, Olson1999}. However, the inescapable difficulty with any numerical model is that the wide range of dynamical length scales present in natural flows simply cannot be accommodated, even on present-day supercomputers. Instead, numerical models employ (either explicitly or implicitly) a fluid viscosity that is typically at least ten orders of magnitude larger than that in astrophysical objects. The use of unrealistically large viscosity 
presents two major difficulties in interpreting computational results. First, the convective vortices assume an artificially large scale in the numerical models; 
more realistic simulations (with lower viscosity) would drive convective flows at much smaller scales.
At these small convective scales,
the magnetic Reynolds number, $\Rm$, the ratio of the Ohmic diffusion timescale
to the magnetic induction timescale, is less than unity for planetary conditions;
realistic small-scale convective flows are therefore unable to maintain magnetic fields
through dynamo action \citep{Jones2000b}.
Second, the simulated flows are considerably less turbulent than those that occur naturally. In the models, convective vortices can produce large-scale magnetic fields in only a relatively laminar regime, where the buoyancy driving is moderate \citep[][]{Christensen2006, Tilgner12}. When 
the driving is increased, although the convective vortices retain their axial structure, they lose their spatial and temporal coherence, thereby diminishing the electromotive force responsible for maintaining the large-scale magnetic field \citep{Cattaneo06, Soderlund2012}. In this case, provided that the magnetic diffusion is sufficiently small, it is small-scale fields (\ie of size comparable with or smaller than the perpendicular lengthscale of the convective vortices) that are generated.

\begin{figure}
	\centering
       \includegraphics[clip=true,width=8.6cm]{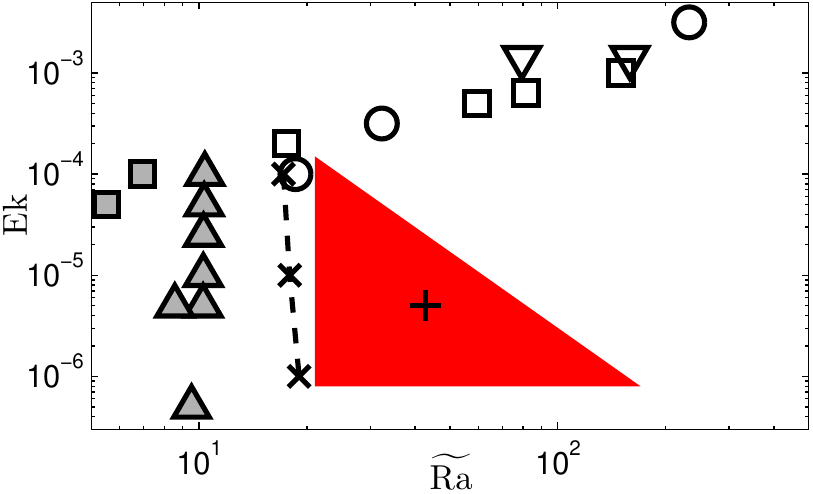}
       \caption{\label{fig:diagram} (color online)
Parameter values in \mbox{$(\Ek,\tRa)$} parameter space for previous studies of convective planar dynamos.  
For compressible convection: \citet{Kapyla2009} ($\Box$) and \citet{Favier2013} ($\bigcirc$). For Boussinesq convection: \citet{Stellmach2004} ($\triangle$), \citet{Cattaneo06} ($\bigtriangledown$) and \citet{Tilgner12} ($\times$). Gray (open) symbols indicate dynamos producing large-scale (small-scale) magnetic fields 
for \mbox{$Pm=\mathcal{O}(1)$}. 
The crosses and the dashed line indicate the transition between the two types of dynamos identified in \citet{Tilgner12}. The red region represents where LSVs occur in non-magnetic convection for aspect ratio \mbox{$\lambda=1$} and \mbox{$\Pran=1$} \citep{Guervilly2014}; the plus symbol is the case studied here. }
\end{figure}

This second point is emphasized in figure~\ref{fig:diagram}, which pinpoints the location in parameter space 
of previous numerical models of convective dynamos in planar geometry 
\citep{Stellmach2004, Cattaneo06, Kapyla2009, Tilgner12, Favier2013}. 
The ordinate plots the Ekman number, $\Ek$, the ratio of the rotation period to the viscous diffusion timescale; 
for comparison \mbox{$\Ek\approx10^{-15}$} in the Earth's liquid core. The abscissa denotes the degree of supercriticality 
of the convection expressed by the rescaled Rayleigh number,
\mbox{$\tRa = \Ra \Ek^{4/3}$}, where the Rayleigh number $\Ra$ measures the ratio of buoyancy driving to dissipative effects. Under the Boussinesq approximation 
the onset of convection is given by \mbox{$\tRa\approx 8.7$} as \mbox{$\Ek\to0$} \citep{Chandrasekhar61}. For compressible convection, $\tRa$ is depth-dependent; the values 
shown in figure~\ref{fig:diagram} are those given in the referenced papers \citep{Kapyla2009, Favier2013}. The gray symbols represent dynamos that produce large-scale  
fields, while the open symbols represent small-scale dynamos. The crosses and the dashed line denote the transition between these two types of dynamo reported by \citet{Tilgner12}, which is in agreement with the other studies, both Boussinesq and  compressible. Importantly, \citet{Tilgner12} emphasized that the transition is located well within the rapidly-rotating convection regime defined by hydrodynamic studies. Indeed, the transition is located close to the onset of convection, even as $\Ek$ is decreased towards more realistic values. However, in astrophysical bodies, it is thought that convection is driven well above onset, in a regime where small-scale dynamos are to be expected if we were to extrapolate previous results to small $\Ek$. Consequently, this suggests that a vital ingredient is missing in these models in terms of explaining the generation of large-scale magnetic fields.
This missing ingredient must rely on an inviscid process that leads to the formation of flows at 
large scales, for which $\Rm$ is sufficiently large to support dynamo action, and operates in a turbulent regime.
This process is traditionally thought to originate from strong magnetic feedback forces acting on the flow, leading to a balance in the momentum equation between magnetic, buoyancy (Archimedean) and Coriolis forces (so-called MAC balance). Here, based on recent work on rotating turbulent convection, we propose an alternative view for the formation of large-scale flows that is hydrodynamical (rather than magnetohydrodynamical) in origin.

In previous studies of convective dynamos,
the flow typically consists of small-scale vortices. However, recent work in non-magnetic rotating planar convection
has demonstrated that large-scale coherent flows can form from
turbulent convective vortices,
for both compressible \citep{Chan07, Kapyla11} and Boussinesq fluids \citep{Rubio13, Favier2014, Guervilly2014}. These large-scale flows consist of depth-invariant, concentrated cyclonic vortices, 
which form by the merger of convective thermal plumes
and eventually grow to the size of the computational domain.
Weaker anticyclonic circulations form in their surroundings.
Two conditions are needed for the formation of a large-scale vortex (LSV): rapid rotation and a sufficient level of convection-driven turbulence. These may be quantified as: (i)~the local Rossby number, a measure of the ratio of rotation period to convective turnover timescale, \mbox{$\lesssim 0.1$}; (ii)~\mbox{$\tRa \gtrsim 20$} 
\citep{Guervilly2014}. In figure~\ref{fig:diagram}, the region bounded by these two conditions 
(\ie the parameter window where LSVs occur in planar geometry) is indicated in red. The bottom line of the red window 
corresponds to the smallest Ekman number employed in \citet{Guervilly2014}, but we expect the window to extend 
towards smaller $\Ek$. Since the range of $\tRa$ over which LSVs occur widens as $\Ek$ decreases, we expect that 
LSVs could well be present in rapidly rotating astrophysical objects. 

In this paper we address the important issue of the nature of the dynamo action resulting from convection containing LSVs. In particular, can LSVs produce large-scale magnetic fields? If this is indeed the case then it offers a possible resolution to two long-standing problems. The first concerns the scale of the flows responsible for the dynamo;
the scale of the LSVs is independent of viscosity and therefore does not become extremely small for $Ek \ll 1$. The second addresses whether large-scale field can be produced far above the onset of convection; although the formation of LSVs does need a small Rossby number, crucially it also requires a certain level of turbulence. 

For computational efficiency we employ a local planar model of rotating Boussinesq convection.
The computational domain is three-dimensional and periodic in the horizontal directions.
A vertical temperature difference, \mbox{$\Delta T$}, is imposed across the layer of depth $d$. The aspect ratio of horizontal to vertical box dimensions is denoted by $\aspect$.
The gravitational field is uniform, \mbox{$\vect{g} = - g \vect{e}_z$}. 
The rotation vector is \mbox{$\Omega \vect{e}_z$}. 
The fluid has kinematic viscosity $\nu$, thermal diffusivity $\kappa$, magnetic diffusivity $\eta$, density $\rho$,
thermal expansion coefficient $\alpha$, and magnetic permeability $\mu_0$, all of which are constant. 
Lengths are scaled with $d$, times with \mbox{$1/(2\Omega)$}, temperature with \mbox{$\Delta T$}, and magnetic field with \mbox{$2\Omega d(\rho \mu_0)^{1/2} $}. 
The 
system of dimensionless governing equations is
\begin{eqnarray}
	&&\pdt{\vel} + \vel \bcdot \bnabla \vel + \vect{e}_z \times \vel =
	- \bnabla p
	+ \Ek \nabla^2 \vel 
	+ \frac{\Ra \Ek^2}{\Pran} \theta \vect{e}_z 
	+ \pleft \bnabla \times \B \pright \times \B ,
	\label{eq:u}
	\\
	&& \pdt{\theta} + \vel \bcdot  \bnabla \theta - u_z = \frac{\Ek}{\Pran} \nabla^2 \theta ,
	\label{eq:theta}
	\\
	&& \pdt{\B} = \bnabla \times \pleft \vel \times \B \pright + \frac{\Ek}{\Pm} \nabla^2 \B ,
	\label{eq:B}
\end{eqnarray}
where \mbox{$\vel=(u_x,u_y,u_z)$} is the (solenoidal) velocity field, $p$ the pressure, $\theta$ the temperature perturbation 
relative to a linear background profile, and \mbox{$\B=(B_x,B_y,B_z)$} the magnetic field. 
The dimensionless parameters are the Rayleigh number, \mbox{$\Ra = \alpha g \Delta T d^3/\kappa \nu$},
the Ekman number, \mbox{$\Ek = \nu/2\Omega d^2$},
and the thermal and magnetic Prandtl numbers, \mbox{$\Pran = \nu/\kappa$} and \mbox{$\Pm = \nu/\eta$}.
The upper and lower boundaries are taken to be perfect thermal and electrical conductors,
impermeable and stress-free. Equations~(\ref{eq:u}) -- (\ref{eq:B}) are solved using a pseudospectral code described in detail in \citep{Cattaneo03}. 

\begin{figure}
	\centering
       \subfigure[]{\label{fig:bx_Pm02}
       \includegraphics[clip=true,width=4.1cm]{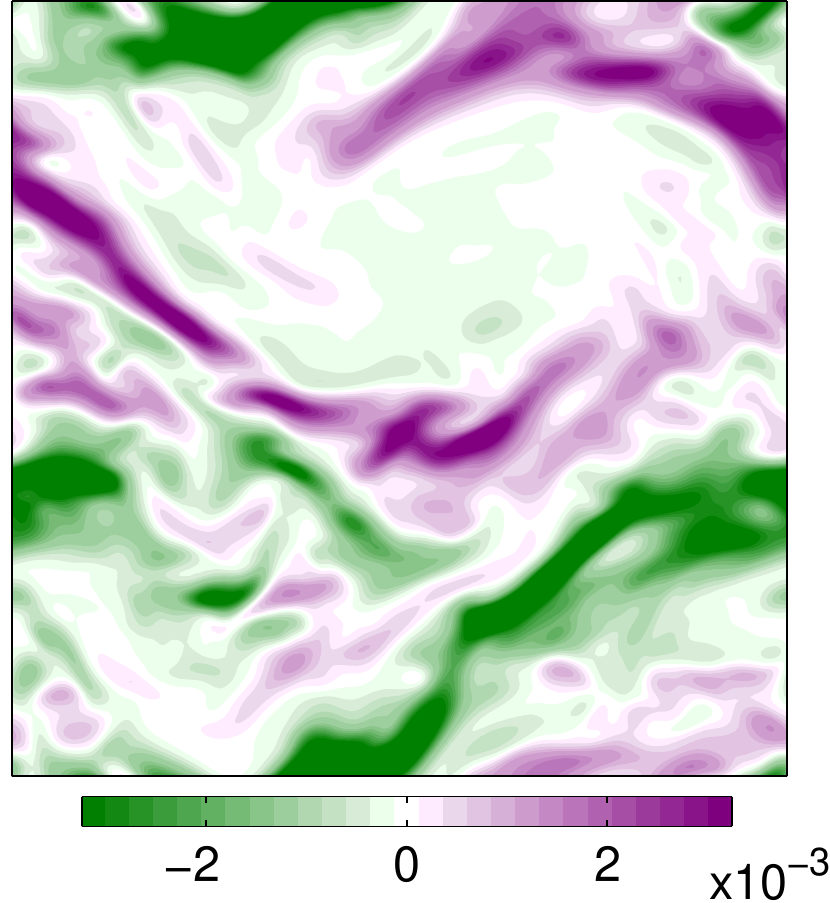}}
       \subfigure[]{\label{fig:bx_Pm25}
       \includegraphics[clip=true,width=4.1cm]{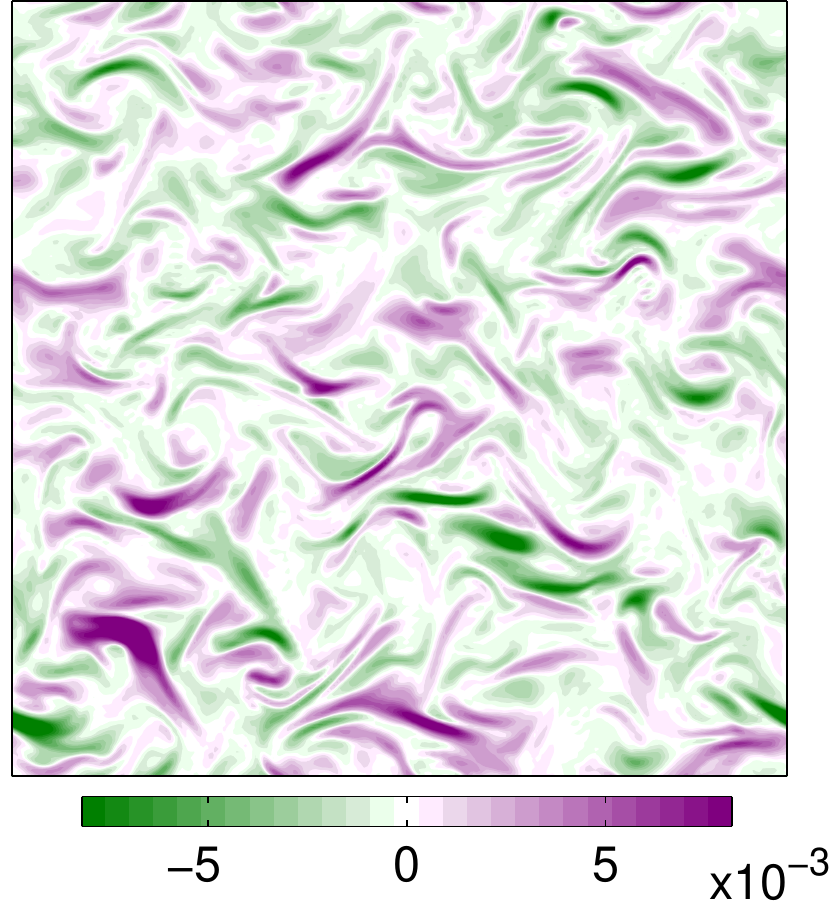}}
       \caption{\label{fig:bx} (color online)
	Horizontal cross-sections of $B_x$ at \mbox{$z=0.25$} for (a) $\Pm=0.2$ and (b) $\Pm=2.5$.}
\end{figure}

We focus on one particular simulation that produces an LSV in the non-magnetic case,
with \mbox{$\Ek=5\times10^{-6}$}, \mbox{$\Pran=1$}, \mbox{$\Ra=5\times10^8$} and \mbox{$\aspect=1$} 
(the plus symbol in figure~\ref{fig:diagram}); the numerical resolution is $256 \times 256 \times 257$ collocation points. 
The Reynolds number, here defined by \mbox{$\Rey=wd/\nu$}, where $w$ is the r.m.s.\ vertical (\ie convective) velocity, is $765$.
We vary only the magnetic Prandtl number $\Pm$, which controls the magnetic diffusivity in equation~(\ref{eq:B}), 
and hence the dynamo threshold. Our results show that coherent large-scale magnetic fields are indeed generated 
in the presence of LSVs, but that the value of $\Pm$ has a crucial influence on the structure of the field sustained by 
the flow. At \mbox{$\Pm=0.2$}, just above the dynamo threshold, the horizontal magnetic field clearly displays 
a system-size structure (figure~\ref{fig:bx_Pm02}), whereas for \mbox{$\Pm=2.5$} the form of the field is drastically 
different, with structures only at much smaller scales (figure~\ref{fig:bx_Pm25}). 

\begin{figure}
	\centering
       \includegraphics[clip=true,width=8.6cm]{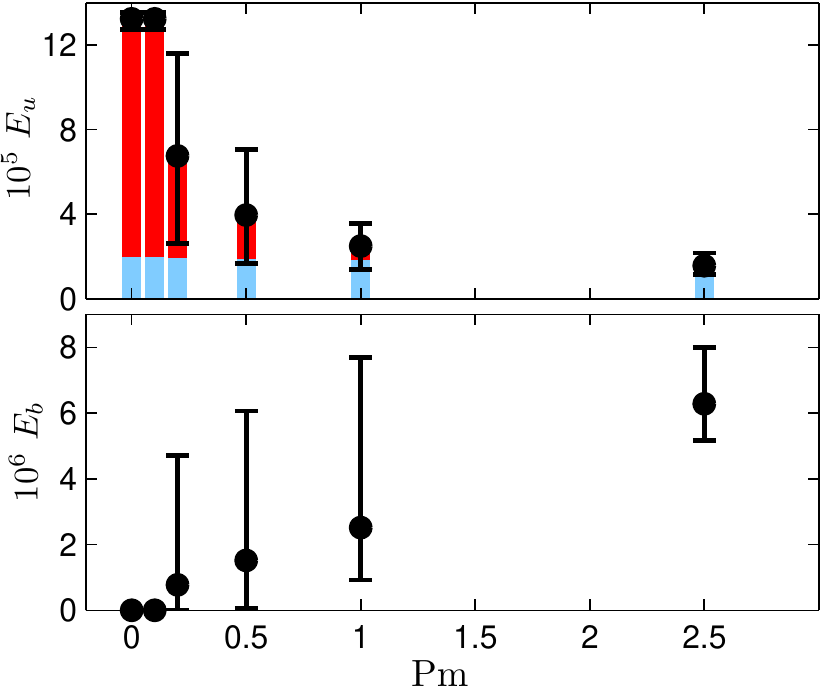}
       \caption{\label{fig:E_Pm} (color online)
Kinetic ($E_u$) and magnetic ($E_b$) energies as a function of $Pm$. The dots indicate the mean values, the black vertical lines the range of variation of the energies in the saturated phase. The red vertical bars show the mean kinetic energy in the LSV (\ie the energy corresponding to the horizontal wavenumber \mbox{$(k_x, k_y) = (1, 1)$)}, the blue bars the energy in the remainder of the flow.}
\end{figure}

\begin{figure}
	\centering
       \subfigure[]{\label{fig:wz_Pm0}
       \includegraphics[clip=true,width=4.1cm]{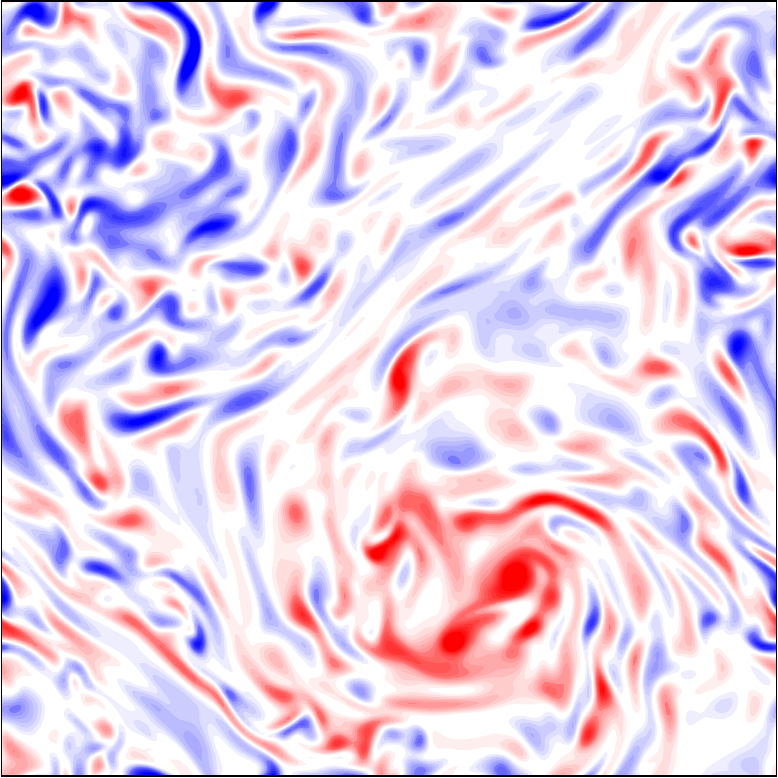}}
       \subfigure[]{\label{fig:wz_Pm25}
       \includegraphics[clip=true,width=4.1cm]{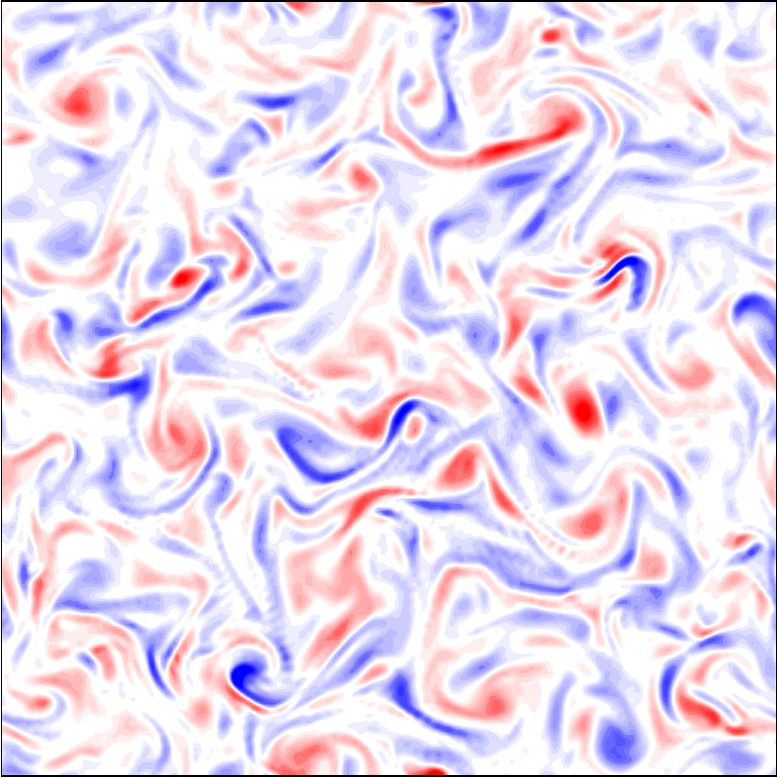}}
       \\
       \subfigure[]{\label{fig:wz_Pm02_max}
       \includegraphics[clip=true,width=4.1cm]{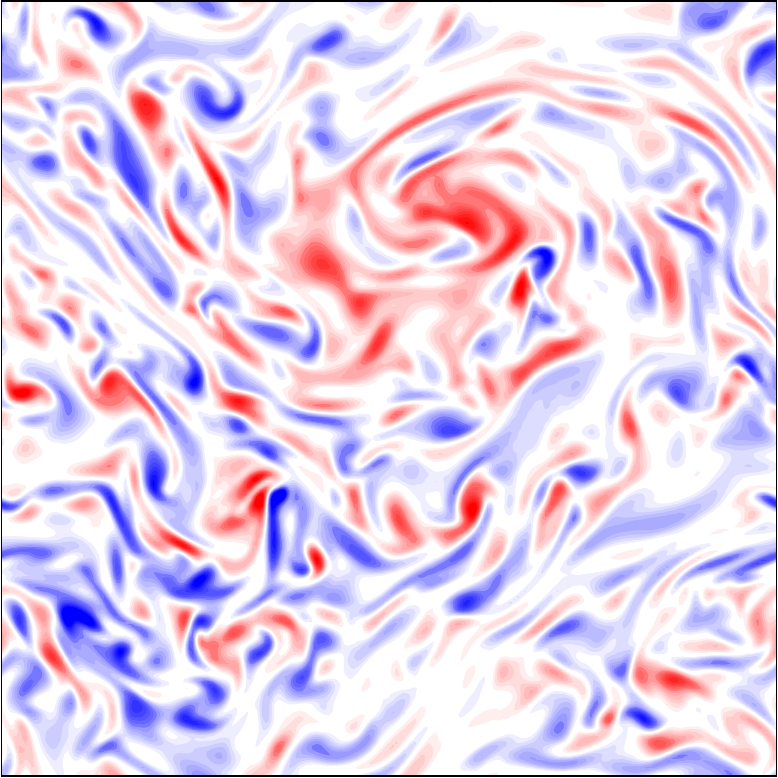}}
       \subfigure[]{\label{fig:wz_Pm02_min}
       \includegraphics[clip=true,width=4.1cm]{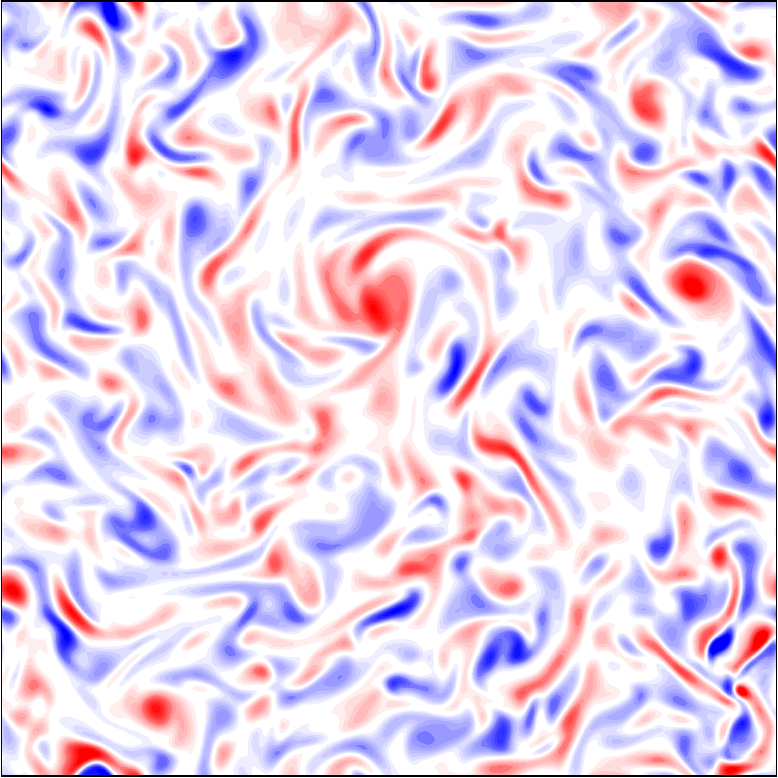}}
       \caption{\label{fig:wz} (color online)
Horizontal cross-sections of the axial vorticity at \mbox{$z=0.25$} for (a) $\Pm=0$, (b) $\Pm=2.5$,
and (c) $\Pm=0.2$ at $t_{\textit{max}}$ and (d) $t_{\textit{min}}$. 
The colorscale is bounded by \mbox{$\pm0.8$} in all the cases.}
\end{figure}

\begin{figure}
	\centering
       \subfigure[]{\label{fig:spectrumu}
       \includegraphics[clip=true,width=4.1cm]{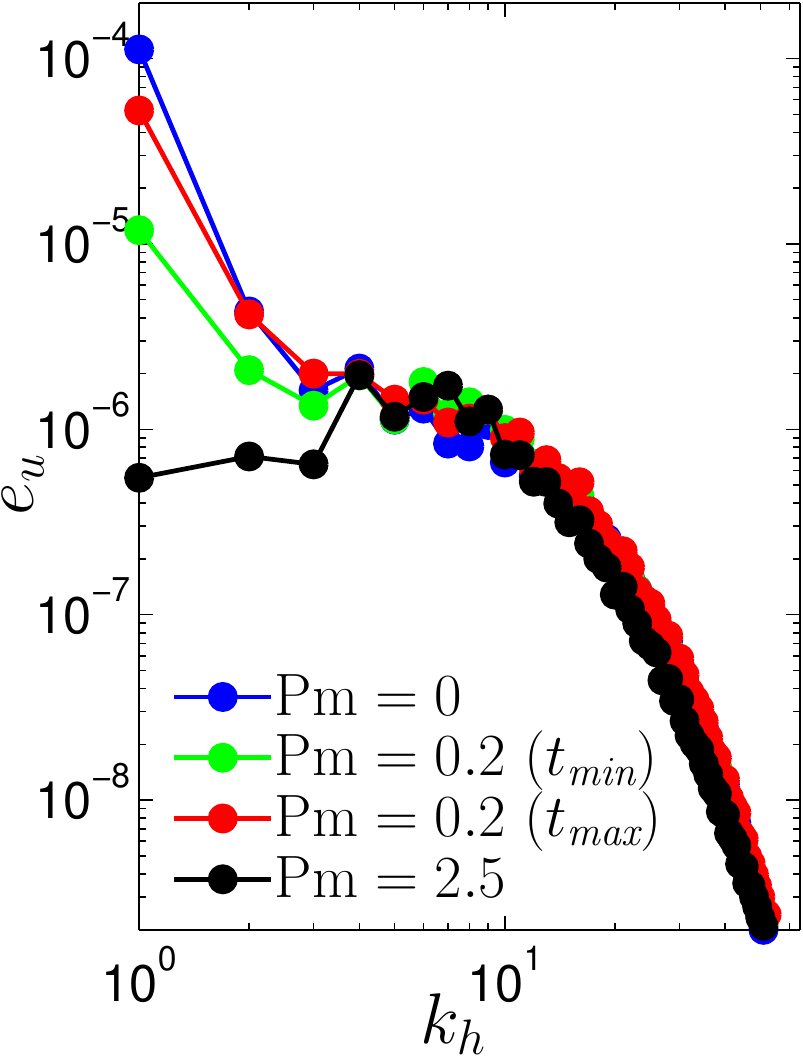} }
       \subfigure[]{\label{fig:spectrumb}
       \includegraphics[clip=true,width=4.1cm]{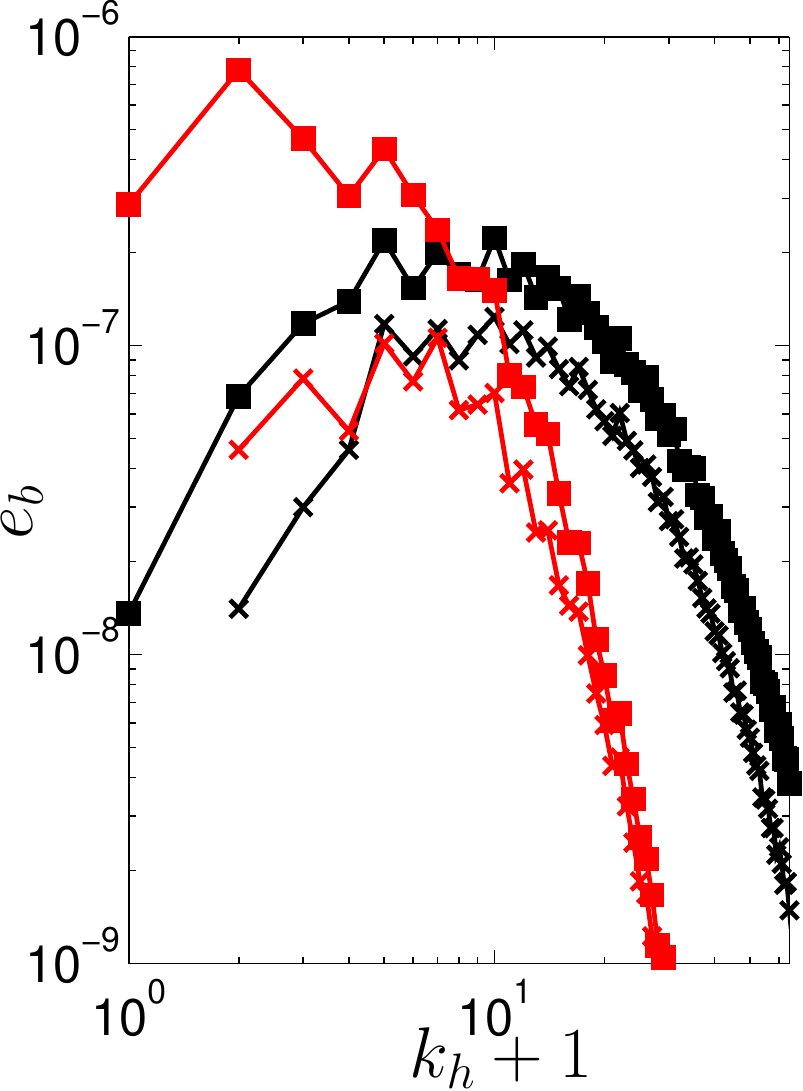}}
       \caption{(color online) Horizontal power spectra of (a) the velocity ($e_u$) and (b) the magnetic 
       field ($e_b$).
       In (b), the squares and crosses represent the horizontal and vertical components of the
       magnetic field respectively, for \mbox{$\Pm=0.2$} at 
       \mbox{$t=t_{\textit{max}}$} (red symbols) and \mbox{$\Pm=2.5$} (black). 
}
\end{figure}

Figure~\ref{fig:E_Pm} shows the mean values of the kinetic and magnetic energies in the saturated (\ie dynamic) phase as a function of $\Pm$ as it varies from zero (the hydrodynamic case) to $\Pm = 2.5$. For this Reynolds number, dynamo action ensues when \mbox{$\Pm\gtrsim0.2$}, \ie for
\mbox{$\Rm\gtrsim 153$}, where \mbox{$\Rm=\Rey \Pm$} is the magnetic Reynolds number. While the magnetic energy increases with $\Pm$, as expected, the kinetic energy undergoes a significant decrease, such that for \mbox{$\Pm=2.5$} the kinetic energy is one order of magnitude smaller than for the hydrodynamic case.  
Above the dynamo threshold, both energies display large fluctuations (indicated by the black vertical lines); these decrease for \mbox{$\Pm=2.5$}, especially those of the kinetic energy. 
The decrease of the total kinetic energy corresponds essentially to the decay of the 
energy of the LSV (represented by the red vertical bars), while the vigor of the convective flows at smaller scales remains relatively unchanged as $\Pm$ is increased (blue bars). The suppression of the LSV when \mbox{$\Pm \gtrsim 1$} is confirmed in figure~\ref{fig:wz_Pm25}, which shows a horizontal cross-section of the axial vorticity 
for \mbox{$\Pm=2.5$}. The flow is dominated by small-scale convection, in sharp contrast with the hydrodynamic case (figure~\ref{fig:wz_Pm0}), in which the flow is organized into a concentrated cyclone at large scale.
The horizontal power spectra of the velocity corresponding
to the snapshots of figure~\ref{fig:wz} are shown in figure~\ref{fig:spectrumu};
the horizontal wavenumber $k_h = n$ includes all modes in the range 
\mbox{$n-1/2 \leq \pleft k_x^2 + k_y^2 \pright^{1/2} < n+1/2$}; 
\mbox{$k_h=1$ corresponds to the LSV}.
The upscale kinetic energy transfer is clearly halted
for \mbox{$\Pm=2.5$} compared with the hydrodynamic case,
while the kinetic energy of the scales in the neighbourhood of or smaller than the dominant convective scale
(\mbox{$k_h\approx10$}) remains unchanged.

The magnetic field generated at \mbox{$\Pm=2.5$} is dominated by small scales (figure~\ref{fig:bx_Pm25}). No coherent field is produced with either a vertical or horizontal large-scale structure. The small-scale field suppresses the LSV, even though the magnetic energy is less than half the kinetic energy. For sufficiently large $\Pm$ (\ie sufficiently large $\Rm$), the suppression of the LSV is an example of small-scale magnetic field impeding the transport properties of the flow, as seen in studies of the interactions between turbulent flows and imposed magnetic fields \citep{Cattaneo1991, Tobias07}.

\begin{figure}
	\centering
       \includegraphics[clip=true,width=8.6cm]{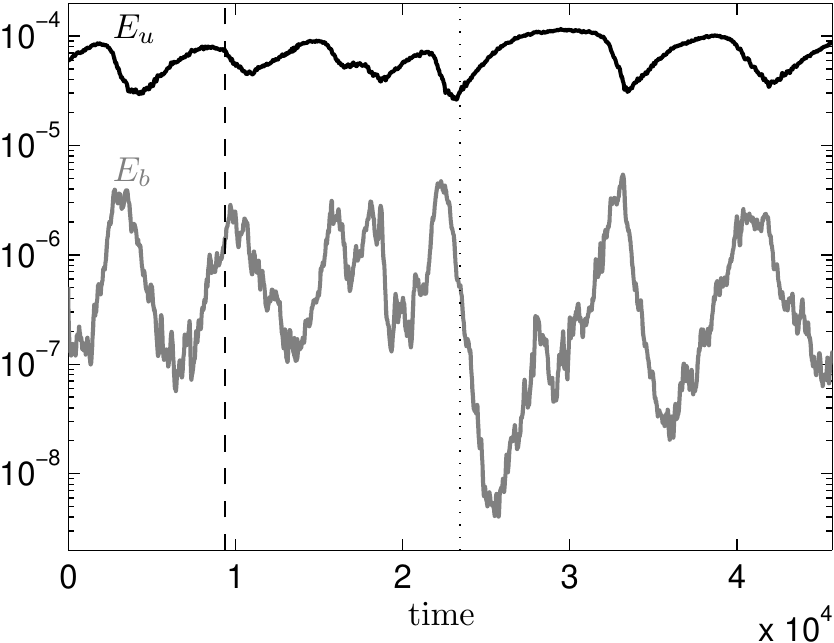}
       \caption{\label{fig:fluc}
Time series of the kinetic ($E_u$, solid black line) and magnetic ($E_b$, gray) energies for \mbox{$\Pm=0.2$}. 
The vertical dashed (dotted) line indicates the time \mbox{$t=t_{max}$} (\mbox{$t=t_{min}$}).}
\end{figure}

By contrast, the behavior at \mbox{$\Pm=0.2$} is very different: figures~\ref{fig:wz_Pm02_max} and \ref{fig:wz_Pm02_min} show two snapshots of the axial vorticity at a maximum (\mbox{$t=t_{\textit{max}}$}) and a minimum (\mbox{$t=t_{\textit{min}}$}) of the kinetic energy in the saturated phase. A large-scale cyclone similar to that produced in the hydrodynamic case is present at \mbox{$t_{\textit{max}}$}. At \mbox{$t_{\textit{min}}$}, the cyclone is significantly weakened and reduced in size, but crucially is not entirely destroyed, unlike for \mbox{$\Pm=2.5$}. 
The power spectra of the velocity (figure~\ref{fig:spectrumu}) show that 
the kinetic energy of the LSV for $ \mbox{$\Pm=0.2$}$ at \mbox{$t_{\textit{max}}$}
is slightly smaller than in the hydrodynamic case. 
Only the amplitude of the largest scales ($1\leq k_h\leq 3$)
varies significantly during the fluctuations of the kinetic energy.
Figure~\ref{fig:fluc} shows the time series of the kinetic and magnetic energies for $\Pm=0.2$. The kinetic and magnetic fluctuations are anti-correlated and correspond to cycles of regeneration and suppression of the LSV. When the magnetic field is weak, the amplitude of the LSV grows, yielding a rapid increase in the magnetic energy; when the field becomes sufficiently strong, the LSV is suppressed, thereby leading to the eventual decrease of the magnetic field. 

The snapshot of $B_x$ in figure~\ref{fig:bx_Pm02} is taken at \mbox{$t=t_{\textit{max}}$}, during a growing phase of the magnetic energy. 
Bands of strong horizontal magnetic field are localized in the shear layers surrounding the LSV, while the magnetic field is weaker in the core of the cyclone. The horizontal power spectra of the horizontal and vertical magnetic field for \mbox{$\Pm=0.2$} at \mbox{$t=t_{\textit{max}}$} are shown in figure~\ref{fig:spectrumb};
\mbox{$k_h=0$} corresponds to the horizontally-averaged mode. The horizontal magnetic energy is dominated by the largest horizontal scales. 
The vertical field 
displays a finer structure dominated by horizontal lengthscales around \mbox{$k_h = 6$}, with only a weak amplitude at larger scales. By comparison, the horizontal and vertical components of the field for \mbox{$\Pm=2.5$} 
are dominated by scales in the neighbourhood of the convective scale, with a peak around \mbox{$k_h=10$}. For \mbox{$\Pm=0.2$}, the LSV never entirely disappears; the magnetic field is then dominated by large-scale bands of horizontal field during the entire cycle.

In order to determine if a small-scale dynamo may be present for \mbox{$\Pm=0.2$}, we reduce the aspect ratio of the box to \mbox{$\aspect=0.25$}, keeping all other parameters constant. For this small aspect ratio, the hydrodynamic convection is dominated by small-scale flows, with no LSV. Here we find that the dynamo threshold is increased to \mbox{$Pm=1$}; thus there is no small-scale dynamo driven by the convective flows for \mbox{$\Pm<1$}. Consequently, the dynamo action observed for \mbox{$0.2\leq \Pm< 1$} with \mbox{$\aspect=1$} relies crucially on the presence of the LSV. The fluctuations of these dynamos are caused by the amplification of the small-scale magnetic field due to interactions between the convective flows and the large-scale magnetic field. This amplification quenches the LSV, which, in turn, leads to a decrease of the magnetic field at all scales. The LSV can eventually regenerate once the small-scale magnetic field has become sufficiently weak.

In summary, we have proposed a new self-consistent dynamo mechanism to explain the generation of system-size magnetic fields by turbulent rotating convection. The dynamo involves two steps: the formation of LSVs from small-scale convective flows, and the generation of large-scale magnetic fields by the action of the LSVs. The large-scale fields concentrate in the shear layers surrounding the LSVs and are essentially horizontal. 
The dynamo operates for small  
$\Pm$
(\ie moderate 
$\Rm$ in our simulations, where $\Rey$ is roughly constant), below the threshold for small-scale dynamo action. 
The competition between the generation of large-scale magnetic fields in the presence of LSVs, which leads to the amplification of small-scale magnetic field by the convective flows, and the subsequent suppression of the LSVs by these small-scale fields, yields the fluctuating behavior of this dynamo. 
Above the small-scale dynamo threshold, the small-scale magnetic field acts on the convective flows so as to prevent the formation of LSVs. 
In our numerical model, at \mbox{$\Ek=5\times 10^{-6}$}, this threshold occurs for \mbox{$\Pm \geq 1$}, \ie
\mbox{$\Rm\geq 765$}. However, the threshold presumably depends on the convective scale
and therefore on $\Ek$, so
in the limit of small $\Ek$, the relevant regime for many astrophysical objects, 
we might expect that
small-scale dynamo action will require much larger $\Rm$.
The suppression of the LSVs by small-scale magnetic fields at high $\Rm$ 
probably restricts the relevance of the novel dynamo mechanism that we propose here
to those astrophysical objects with moderate $\Rm$. 
Significantly, however, this includes both terrestrial and gas planets, for which 
\mbox{$\Rm=\mathcal{O}(10^3-10^5)$} at the system size and \mbox{$\Rm<1$} at the convective scale.
The importance of this dynamo 
is its robustness in the limit of small Ekman numbers. Indeed, unlike traditional self-consistent convective dynamo 
models that rely on the presence of coherent, viscously-controlled columnar flows, the dynamo discussed here relies on 
the formation of LSVs, which is controlled by nonlinear inviscid energy transfers in rapidly-rotating systems. It is important 
to note that the LSVs consist essentially of horizontal flows, so they cannot themselves act as dynamos \citep{Zeldovich1957}. 
In a forthcoming study, we shall investigate the dynamo mechanism in detail, and, in particular, describe how the LSVs modify 
the three-dimensional flows so as to generate large-scale magnetic fields.

\vspace{11pt}
\emph{Acknowledgments} 
This work was supported by the Natural Environment Research Council under grant NE/J007080/1. This work was undertaken on ARC1 and ARC2, part of the High Performance Computing facilities at the University of Leeds. This work also used the COSMA Data Centric system at Durham University, operated by the Institute for Computational Cosmology on behalf of the STFC DiRAC HPC Facility (www.dirac.ac.uk). This equipment was funded by a BIS National E-infrastructure capital grant ST/K00042X/1, DiRAC Operations grant ST/K003267/1 and Durham University. DiRAC is part of the National E-Infrastructure.
We are grateful to three anonymous referees for helpful suggestions.

\end{document}